\documentclass[journal]{IEEEtran}
\usepackage{ulem}
\usepackage{textcomp}

\usepackage{cite}

   \usepackage{graphicx}

\ifCLASSINFOpdf
   \usepackage[pdftex]{graphicx}
\else
\fi
\usepackage[cmex10]{amsmath}

\begin{document}

\title{Robust design by anti-optimization for parameter tolerant GaAs/AlOx High Contrast Grating mirror for VCSEL application }


\author{Christyves~Chevallier,
        Fr\'ed\'eric~Genty,
        Nicolas~Fressengeas,
        and~Jo\"el~Jacquet
\thanks{This work was supported by the French ANR in the framework of Marsupilami project (ANR-09-BLAN-0166-03)
and by INRIA and R\'egion Lorraine with the InterCell grant (http://intercell.metz.supelec.fr - CPER2007).}
\thanks{C. Chevallier, F. Genty and J. Jacquet are with Sup\'elec, Laboratoire Mat\'eriaux Optiques,
Photonique et Syst\`emes, EA 4423, 2, rue E. Belin 57070 Metz, France. E-mail: christyves.chevallier@supelec.fr}
\thanks{N. Fressengeas is with Universit\'e de Lorraine, Laboratoire Mat\'eriaux Optiques,
Photonique et Syst\`emes, EA 4423, 2, rue E. Belin 57070 Metz, France.}
\thanks{Copyright (c) 2013 IEEE. Personal use of this material is permitted. 
However, permission to use this material for any other purposes must be obtained from the IEEE by sending a request to pubs-permissions@ieee.org.}}

\markboth{Journal of Lightwave Technology ,~Vol.~31, No.~21, Novembre~2013}%
{Chevallier \MakeLowercase{\textit{et al.}}}

\maketitle

\begin{abstract}A GaAs/AlO$_x$ high contrast grating structure design which exhibits a 99.5~\% high reflectivity
for a 425~nm large bandwidth is reported. The HCG structure has been designed 
in order to enhance the properties of mid-infrared VCSEL devices by replacing the top Bragg mirror of the cavity.
A robust optimization algorithm has been implemented to design the high contrast grating structure not only
as an efficient mirror but also as a robust structure against the imperfections of fabrication. The design
method presented here can be easily adapted for other high contrast grating applications at different wavelengths.
\end{abstract}


%
\IEEEpeerreviewmaketitle
\section{Introduction}
\IEEEPARstart{H}{igh} contrast gratings (HCG) are diffractive structures made of a material with a high optical
index for the grating slabs surrounded by a low index material. With a high optical index contrast ($\sim2$) and
a near-wavelength grating period, the structure diffracts only into the 0$^{\mbox{th}}$ order of diffraction and
can be seen as a 1D photonic crystal supporting only a few propagative Bloch modes. However, contrary to photonic
crystal slabs, the modes propagate perpendicularly to the slab plane\cite{lalanne_jlt_2006,changhasnain_aop_2012}.
The resonance of these modes between the two grating interfaces and their coupling at the interfaces can be adjusted
to obtain different and very promising properties in a large range of applications\cite{changhasnain_aop_2012} such
as broadband mirrors\cite{goeman_ptl_1998,mateus_ptl_2004_2,wu_oe_2012,foley_ol_2012}, high-Q resonators\cite{zhou_oe_2008},
planar lenses\cite{fattal_np_2010}, wavefront control \cite{carletti_oe_2011}, optical isolators\cite{ye_oe_2010},
 waveplates\cite{magnusson_ol_2010}, circular polarizers\cite{mutlu_ol_2012} or electromechanical 
 mirrors\cite{hane_apl_2006,huang_np_2008}. The mirror effect is particularly interesting since it can exhibit not only
 a polarization independence\cite{shokooh-Saremi_ol_2010} with good angular insensitivity\cite{wu_jopt_2010_b}  
 but also a good reflectivity selectivity between TE and TM  modes\cite{goeman_ptl_1998,shokooh-saremi_oic_2010}.
 With only one high contrast grating layer, it is thus possible to obtain a very large bandwidth of $\Delta\lambda/\lambda =$ 30\%
 for more than 99\% reflectivity and even $\Delta\lambda/\lambda =$ 17\% for 99.9\%\cite{mateus_ptl_2004}.

  The development of VCSEL devices emitting in the 2\nobreakdash--3~\textmu m wavelength range remains a challenging
  task today. The realization of VCSEL devices which are single-mode, low cost and tunable light sources \cite{vcsels_ch10}
  is of great interest for gas detection in the mid-infrared wavelength range where gas species such as carbon monoxide
  exhibit strong absorption lines\cite{ouvrard_ptl_2005,chen_apb_2010}. 
  VCSEL structures based on InP system have demonstrated mid infrared laser emission in continuous wave (CW) at room
  temperature for $\lambda = $ 2.3 \textmu m \cite{ortsiefer_el_2006}. However this wavelength seems to be the maximum
  limit for this material system \cite{vcsels_ch10}. Thus several VCSEL structures based on AlGaInAsSb material system
  have been developed. A VCSEL structure using a buried tunnel junction for current confinement have shown CW emission
  at $\lambda = $~2.36~\textmu m up to 363~K\cite{bachmann_pp_2010} while another structure based on selective lateral
  etching of the tunnel junction in order to realize a current aperture have shown CW emission at $\lambda =$~2.31~\textmu m up to 343~K\cite{sanchez_oe_2012}.
  More recently, a GaSb-based VCSEL using a lateral wet oxidation of AlAs for current confinement has been demonstrated and exhibits
  CW laser emission at $\lambda  =$~2.38~\textmu m for a temperature of 253~K\cite{laaroussi_el_2012}. The development of VCSEL based on AlGaInAsSb is 
  thus a very promising solution for the mid infrared wavelength range. However, laser emission is still limited today at
  $\lambda = $~2.6~\textmu m at room temperature \cite{arafin_apl_2009,ducanchez_el_2009_2}.
  One of the main problem at such large wavelengths is the increase of the device thickness which reaches about 12 \textmu m since
  more than 20 pairs of quarter-wavelength AlSb/GaSb layers are required for the VCSEL Bragg mirrors\cite{cerutti_jcg_2009}. 

  The large and high reflectivity bandwidth, low mirror thickness and high polarization selectivity of HCG structures make them 
  good candidates for replacement of Bragg mirrors in VCSEL diodes.
  VCSEL designs which use a grating as a polarizing mirror in order to enhance the laser properties have thus been
  proposed~\cite{magnussson_patent_2000,magnusson_spie_2003}. 
  Several VCSEL structures based on HCG have then been developed and have demonstrated laser emission with GaAs-based
  material around 850 nm\cite{huang_np_2007} and 980 nm\cite{gilet_spie_2010} or with InP-based material at 1320 nm\cite{hofmann_pj_2010}
  and 1550 nm \cite{boutami_apl_2007,chase_oe_2010,sciancalepore_ptl_2012}.

  The short amplification length of VCSEL structures imposes the use of a high quality mirror with reflectivity
  larger than 99.5~\% for large bandwidths which are typically about 150~nm in mid-infrared\cite{cerutti_jcg_2009}.
  The mirror efficiency required for VCSEL application is thus demanding and imposes a precise adjustment of HCG
  geometrical parameters during the design process. Even if the physic of HCG is well understood~\cite{lalanne_jlt_2006,karagodsky_oe_2010},
  numerical simulations by RCWA~\cite{moharam_josaa_1995} of the reflectivity of HCG structures allow the use of an
  optimization algorithm~\cite{bisaillon_oe_2006,shokooh-saremi_ol_2007,wu_jopt_2010_a,chevallier_apa_2010} to design
  the mirror properties for a specific application. 

  Since VCSEL devices require high quality mirrors and HCG have sub-micrometric square-shaped patterns,
  the fabrication of HCG structures needs to control masking and etching process with a high accuracy in
  the nanometric range\cite{chevallier_apa_2010}. On the other hand, HCG mirrors can be designed to have
  a good robustness of several percent on the geometrical dimensions. So, to achieve a robust and efficient mirror,
  the tolerance with respect to the errors of fabrication has to be taken into account during
  the design of the structure\cite{chevallier_olt_2011,zhou_ptl_2008,wu_jopt_2010_a}.

  In this paper, we present the design of a GaAs HCG combined with an AlO$_x$ sublayer as low index material to replace
  the top DBR of mid infrared GaSb-based VCSEL. In a first part, an optimization algorithm is used to find the best
  dimensions of the HCG structure. Then, the tolerance of the geometrical parameters of the optimum design with respect
  to the errors of fabrication are numerically investigated. In a second part, an anti-optimization algorithm is combined
  to the optimization process to develop a robust optimization algorithm. This original design approach of high contrast
  gratings takes into account the tolerances required by the manufacturers on the different parameters directly during
  the design process. The gratings mirrors are thus optimized to exhibit not only high efficiency but also large tolerance values.

\section{Optimization method}
  \label{part_Optim}

  \subsection{Structure of the mirror}

    \begin{figure}[!t]
      \centering
      \includegraphics[width=2in]{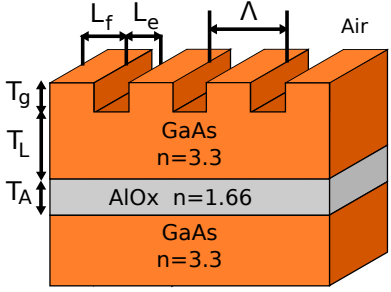}
      \caption{Scheme of the GaAs/AlO$_x$ mirror. The grating period $\Lambda$, Fill Factor $FF = L_f/\Lambda$,
      grating thickness $T_g$, GaAs layer thickness $T_L$ and AlO$_x$ layer thickness $T_A$ are optimized by an
      optimization algorithm to exhibit reflectivities higher than 99.5~\% for a VCSEL application at 2.3 \textmu m.}
      \label{scheme}
    \end{figure}

    The mirror structure presented in Figure \ref{scheme} is made of a GaAs grating ($n~=~3.3$) with a period
    $\Lambda$, a thickness $T_g$ and a Fill Factor $FF = L_f/\Lambda$. The high contrast of optical index required
    for large bandwidth mirror is obtained thanks to the use of a low index  AlO$_x$ sublayer ($n~=~1.66$) with
    a thickness $T_A$. The choice of the GaAs/AlO$_x$ material system has been made in order to allow a monolithic
    epitaxial process during the integration of the HCG mirror in a VCSEL structure\cite{almuneau_jopt_2011,gilet_spie_2010}.
    The AlO$_x$ layer can be obtained from a wet oxidation process of AlAs similar to the one used for current confinement by
    oxide aperture in VCSEL structures\cite{huffaker_apl_1994,choquette_el_1994,laaroussi_jpd_2011}. The GaAs top layer is not
    completely etched during the grating fabrication resulting in a GaAs intermediate layer of thickness $T_L$. The presence of
    an intermediate GaAs layer should enhance the mechanical stability of the grating above the AlO$_x$ layer since 
    the oxidation process decreases the AlAs layer thickness of about 10~\%\cite{almuneau_jopt_2011}.

    HCG mirrors presented in this work are designed to be used as VCSEL top mirrors, thus in the HCG structure presented
    in Figure~\ref{scheme}, the GaAs substrate corresponds to the VCSEL cavity with light propagating from substrate to
    the air. In order to obtain a laser cavity, a minimum reflectivity value of 99.5~\% is required for VCSEL mirrors for
    the largest possible bandwidth. To solve the problem of polarization instability and mode hoping of VCSEL devices, the
    mirror is chosen to be polarization dependent by reflecting light only for the transverse magnetic mode (TM) while keeping
    the transverse electric (TE) reflection coefficient $R_{TE}$ below a 90~\% threshold.

    \subsection{Optimization of the mirror parameters}

    In order to satisfy all the previously defined VCSEL requirements, the structure dimensions
    $\Lambda$, $T_g$, $FF$, $T_A$ and $T_L$ have to be carefully designed. An optimization algorithm
    has been used to automate the search of the most efficient mirror design. The efficiency of the HCG
    structure has been defined through the use of a figure of merit $MF$\cite{chevallier_apa_2010} which
    represents quantitatively the mirror quality from a VCSEL application point of view :

    \begin{equation}
      \label{eq_q}
      MF = \frac{\Delta\lambda}{\lambda_0}\frac{1}{N}\sum_{\lambda=\lambda_1}^{\lambda_2}{R_{TM}(\lambda)g(\lambda)}
    \end{equation}
     
    The figure of merit $MF$ mainly represents the normalized bandwidth of the mirror, defined 
    as the wavelength range $\Delta\lambda=\lambda_2-\lambda_1$ around $\lambda_0$ where the reflectivity
    is larger than 99.5~\% for TM mode and below 90~\% for the TE mode. The normalized bandwidth is also
    multiplied by a Gaussian weighted average of the transverse magnetic reflection coefficients $R_{TM}$
    of the bandwidth to ensure a centering around $\lambda_0$. Reflection spectra of the mirror are computed
    by rigorous coupled wave analysis (RCWA)\cite{camfr} for transverse magnetic and transverse electric polarizations.

    Several optimization algorithms can be used to design HCG structures such as simulated
    annealing\cite{bisaillon_oe_2006}, genetic algorithm\cite{shokooh-saremi_ol_2007} or 
    particle swarm\cite{shokooh-Saremi_ol_2010,wu_jopt_2010_a}. Due to the presence of many local maxima,
    a global optimization approach is required. In this work, we have implemented a 
    particle swarm optimization (PSO) algorithm\cite{kennedy_nn_1995} to maximize the figure of merit $MF$.
    The particle swarm algorithm is based on a population of particles which are candidate solutions sharing
    their knowledge of optimum positions when exploring the search space. The particles of the swarm, of which
    positions correspond to a set of the design parameters {$X=\{\Lambda$, $T_g$, $FF$, $T_A$, $T_L\}$}, 
    are moved at each algorithm iteration with a velocity $v_{i,p}$ :

    \begin{equation}
     v_{i,p} = v_{i-1,p} + c_l * (x^l_p - x_{i-1,p})+ c_g*(x^g_p - x_{i-1,p})
    \end{equation}

    The velocity $v_{i,p}$ at the iteration $i$ of the particle $p$ of the swarm is composed of 3 terms.
    Firstly, the inertia of the particle is taken into account by keeping the velocity of the previous 
    iteration $v_{i-1,p}$. Secondly, a local velocity term moves the particle toward the local best 
    position $x^l_p$ known by the particle $p$. Finally, the swarm concept is created by sharing the 
    best position of all particles thanks to the global best position $x^g_p$ which creates a global 
    velocity term. The parameters $c_l$ and $c_g$ are two weights for the local and global velocities 
    which are randomly chosen from a uniform law in the range $[0,2]$ at each particle move\cite{kennedy_nn_1995}.

    The dimensions of the mirror structure have been optimized within bounds that take technological 
    constraints into account. The AlO$_x$ layer thickness $T_A$ has been shown to maximize the reflectivity
    for values of $(2k-1)\lambda/4$\cite{almuneau_jopt_2011}, but has been bounded between 300~nm and
    400~nm to limit the optical losses within the oxide. The Fill Factor limitations are chosen with 
    respect to the e-beam lithography and etching process and have been bounded between 35~\% and 55~\%.
    The other parameter constraints are only given to define a search space for the optimization algorithm
    but can also be chosen with respect to other technological limitations : 500~nm $< T_g <$ 1100~nm,
    900~nm $< \Lambda <$ 1300~nm and 50~nm $< T_L <$ 1000~nm.

  \subsection{Optimization results}

    \begin{table}[!t]
    \renewcommand{\arraystretch}{1.2}
    \caption{Optimum HCG dimensions obtained by a differential evolution algorithm and particle swarm optimization homemade algorithm.}
    \label{table_Optim}
    \centering
    \begin{tabular}{lcc}
      \hline
      & Differential & Particle \\
      & Evolution & Swarm \\
      \hline 
      $T_g$ & 685~nm & 657~nm
	  \\
      $FF$ &  0.513 & 0.538
	  \\
      $T_A$ &  390~nm & 352~nm
	  \\
      $\Lambda$ & 1158~nm & 1117~nm
	  \\
      $T_L$ & 268~nm & 283~nm
	  \\
      \hline
      $\lambda_0$ &  2340~nm  & 2298~nm
	   \\
      $\Delta\lambda$ & 481~nm  & 493~nm
	  \\
      $\Delta\lambda/\lambda_0$&  20.6 \%  & 21.5 \%
	  
    \end{tabular}
    \end{table}

\begin{figure*}[!t]
      \centering
      \includegraphics[width=4.5in]{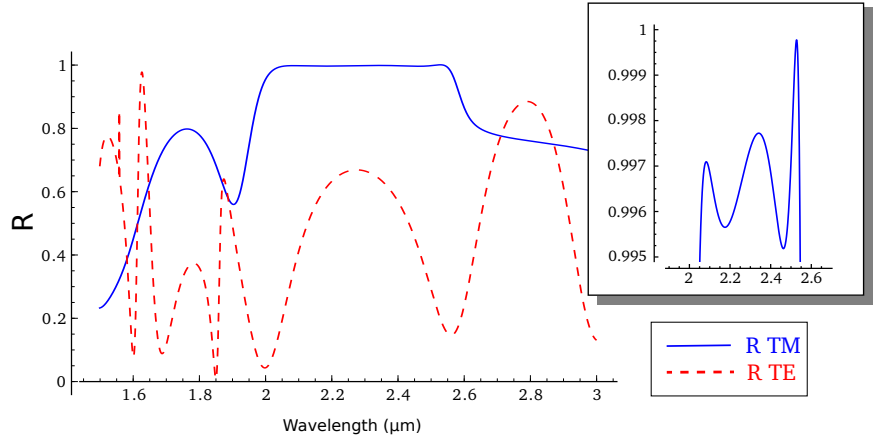}
      \caption{Reflection spectra for transverse magnetic mode (blue) and transverse
      electric mode (dashed red) of the optimum design optimized by the particle swarm
      algorithm which dimensions are described in Table~\ref{table_Optim}. 
      The inset shows the 493~nm large reflectivity bandwidth well centered at $\lambda_0 = $ 2.3 \textmu m.}
      \label{spectre_APSO}
    \end{figure*}

    The particle swarm optimization algorithm has been executed with the previously defined
    parameters to design the GaAs/AlO$_x$ HCG mirror presented in Figure~\ref{scheme}. The
    optimum mirror found by the algorithm exhibits a 493~nm large bandwidth with structure 
    dimensions of $T_g = $~657~nm, $FF= $ 0.5380, $T_A = $~352~nm, $\Lambda = $~1117~nm and
    $T_L = $~283~nm (Table~\ref{table_Optim}). In order to validate the result obtained by
    our implementation of the particle swarm optimization, a comparison has been made with a
    differential evolution\cite{storn_jgo_1997} optimization algorithm\cite{openopt} under
    the same technological constraints for the mirror structure. The execution of the latter
    algorithm results in a 481~nm large bandwidth (Table~\ref{table_Optim}) which is equivalent
    to the PSO algorithm. Both points are very close and satisfy all VCSEL requirements and 
    technological constraints. The 99.5~\% high TM reflectivity and large bandwidth ($\Delta\lambda / \lambda =$ 21.5~\%)
    of the mirror optimized by PSO exhibits a good polarization selectivity by keeping $R_{TE}$
    below 70~\% (Figure~\ref{spectre_APSO}). The high reflectivity performances of 
    the GaAs/AlO$_x$ HCG structure makes it a very promising mirror for VCSEL application at 2300~nm.

  \subsection{Tolerance of the optimum design}

    \begin{table}[!t]
    \renewcommand{\arraystretch}{1.2}
    \caption{Tolerance of the HCG optimized by differential evolution ensuring $R_{TM} > 99.5$ \% and $R_{TE} < $ 90 \% at $\lambda_0$.}
    \label{table_DE}
    \centering
    \begin{tabular}{lllll}
      \hline
      & Optimum & Min & Max & Tolerance  \\
      \hline 
      $T_g$ & 685~nm &
	  682~nm & 756~nm &
	  $\Delta T_g = \pm$3~nm
	  \\
      $FF$ & 0.513 &
	  0.507 & 0.590 &
	  $\Delta FF =  \pm$0.006
	  \\
      $T_A$ & 390~nm &
	  364~nm  &  $>$499~nm &
	  $\Delta T_A = \pm$26~nm
	  \\
      $\Lambda$ & 1158~nm&
	1014~nm & 1163~nm &
	  $\Delta \Lambda = \pm$5~nm
	  \\
      $T_L$ & 268~nm &
	  221~nm & 272~nm &
	   $\Delta T_L = \pm$4~nm\\
      \hline
      $\lambda_0$ &
	  \multicolumn{4}{l}{2340~nm }
	   \\
      $\Delta\lambda$ &
	  \multicolumn{4}{l}{481nm }
	  \\
      $\Delta\lambda/\lambda_0$&
	  \multicolumn{4}{l}{20.6 \% }
	  
    \end{tabular}
    \end{table}

    \begin{table}[!t]
    \renewcommand{\arraystretch}{1.2}
    \caption{Tolerance of the HCG optimized by the particle swarm algorithm ensuring $R_{TM} > 99.5$ \% and $R_{TE} < $ 90 \% at $\lambda_0$.}
    \label{table_APSO}
    \centering
    \begin{tabular}{lllll}
      \hline
      & Optimum & Min & Max & Tolerance  \\
      \hline 
      $T_g$ & 657~nm& 
	  609~nm & 737~nm &
	  $\Delta T_g = \pm$48~nm
	  \\
      $FF$ & 0.538 &
	  0.418 & 0.607 &
	  $\Delta FF =  \pm$0.069
	  \\
      $T_A$ & 352~nm &
	  282~nm & 447~nm &
	  $\Delta T_A = \pm$70~nm
	  \\
      $\Lambda$ &1117~nm &
	993~nm & 1148~nm &
	  $\Delta \Lambda = \pm$31~nm
	  \\
      $T_L$ &283~nm &
	  248~nm & 308~nm &
	   $\Delta T_L = \pm$25~nm\\
      \hline
      $\lambda_0$ &
	  \multicolumn{4}{l}{2298~nm }
	   \\
      $\Delta\lambda$ &
	  \multicolumn{4}{l}{493~nm }
	  \\
      $\Delta\lambda/\lambda_0$&
	  \multicolumn{4}{l}{21.5 \% }
	  
    \end{tabular}
    \end{table}

    From a fabrication point of view, it is important to know how tolerant the structure is with
    respect to the error of fabrication on the different dimensions. The tolerance of one parameter
    is defined as the variation range for which the mirror keeps a TM reflectivity larger than 99.5~\%
    together with a TE reflectivity smaller than 90~\% at $\lambda_0 =$~2300~nm. The evaluation of
    the tolerance of one dimension of the structure is done by increasing and decreasing its optimum
    value while keeping all the other ones at their optimum values. It is important to note that the
    way the evaluation of tolerance is performed does not give any information on an error of fabrication
    made simultaneously on different dimensions\cite{chevallier_olt_2011}.

    As it has already been reported in the literature, HCG can exhibit large tolerances on
    the design parameters\cite{chevallier_olt_2011,zhou_ptl_2008,wu_jopt_2010_a}. For instance,
    minimum and maximum values of the parameters of the optimum designs found by the algorithms 
    result in variation range of more than 10~\% as it can be seen in Table~\ref{table_DE} and 
    Table~\ref{table_APSO}. Despite large variation ranges, the tolerance value of the dimension 
    can be as small as $\pm$~3~nm for instance on the grating thickness $T_g$ (Table~\ref{table_DE}).
    Such a critical tolerance value would make the etching control difficult and decrease drastically
    the probability of successfully etching the grating. Since the optimization algorithm will search
    for the most efficient design regardless of its tolerance, if an optimum is localized at the edge
    of the variation range, the optimization can result in a non tolerant design\cite{chevallier_olt_2011,kontio_ol_2010}.

    The design of GaAs/AlO$_x$ HCG by an optimization algorithm provides very efficient mirrors with
    large bandwidth well adapted for VCSEL integration. However the optimization process can randomly
    result in a point very sensitive to the error of fabrication and can be difficult to fabricate.

\section{Robust optimization}

  \subsection{Anti-optimization method}
    \label{section_AO}
    In order to make the fabrication of HCG possible for a VCSEL integration, mirror structures have
    to be not only efficient but also robust with respect to the fabrication errors. A simple way to
    improve the mirror tolerance is to manually center the structure dimensions within their variation
    ranges\cite{chevallier_olt_2011}. However, since the tolerance is evaluated for each parameter 
    independently, the centering within the variation range does not take into account the error made 
    on several parameters simultaneously. The tolerance computation of combination of parameters such 
    as $L_f = FF*\Lambda$, can increase the knowledge of the robustness, but the adjustment of the grating
    dimensions becomes complex when the number of parameters increases. Moreover, since the manual robustness
    enhancement can only be performed once the dimensions are optimized, the use of an optimization algorithm
    to make the design process easy and automated is lost.

    The maximization of the figure of merit $MF$ in order to find the most efficient mirror structure
    by an optimization algorithm can be coupled to an anti-optimization process to form a robust optimization
    algorithm. The purpose of anti-optimization\cite{elishakoff_cs_1994} is to search for the worst scenario
    of fabrication of the optimum found which would result in the least efficient mirror. For that purpose, a new figure of merit $MF'$ has been introduced : 

    \begin{multline}
    \label{eq_mf2}
      MF'(X) = \frac{1}{2}\left( \underset{\Delta X}{ \mbox{ min }} MF(X \pm \Delta X) + MF(X) \right) (1+\eta) \\
     \forall (X \pm \Delta X) \in \Omega
    \end{multline}

    The robustness of the design is taken into account in the figure of merit $MF'$ for a design
    defined by  $X = \{T_g, FF, T_A, \Lambda, T_L\}$ by computing the average value of the figure
    of merit $MF$ with the minimum of figure of merit associated to the worst scenario of fabrication.
    The search of the minimum is done in the hyper-space $\Omega = X \pm \Delta X$ around $X$ delimited
    by user-defined tolerance specifications with $\Delta X = \{\Delta T_g, \Delta FF, \Delta T_A, \Delta \Lambda, \Delta T_L\}$.
    The last parameter $\eta$ of the figure of merit $MF'$ corresponds to a percentage of the process of tolerance evaluation.
    Thus, efficient mirrors which keep good performances within user-defined variation ranges will have a larger figure of merit $MF'$ than non tolerant structures.

    The evaluation of $MF'(X)$ requires to know the minimum value of $MF(X)$ within the whole tolerance 
    area defined by $X \pm \Delta X$. The search for this minimum could be done by an optimization algorithm 
    but would be time consuming. Moreover, if the current point $MF(X)$ has a low value, the evaluation of its
    robustness would be useless since a weak solution will not be kept during the optimization. Since the
    particle swarm algorithm moves the particles by directing them to local and global optimum values ($x^l_p$ and  $x^g_p$) 
    stored in memory, the tolerance evaluation can be done only on the best positions known by the swarm. In order to make
    a fast evaluation of the tolerance which does not rely on an optimization algorithm to find the worst 
    design within the tolerance area, only the extremum values of each parameter
    $T_g\pm\Delta T_g, FF\pm\Delta FF, T_A\pm\Delta T_A, \Lambda\pm\Delta \Lambda$ and $T_L\pm\Delta T_L$ are computed. 
    The estimation of the minimum used for the averaging of the $MF'$ in 
    Eq.~\ref{eq_mf2} is thus evaluated with only 10 points in the case of the GaAs/AlO$_x$ HCG structure and
    does not take into account errors made simultaneously on several parameters or a local minimum located between
    $T_g$ and $T_g+\Delta T_g$ for instance. 
    In a $N_D$ dimension problem, with $N_D$ parameters to optimize, the tolerance evaluation process is 
    done in $2N_D$ steps for the positive and negative tolerance estimation of each parameter. 
    The parameter $\eta$ in Eq.~\ref{eq_mf2} represents thus the percentage of achievement of the $2 N_D$ 
    tolerance evaluation process. At each step, the figure of merit $MF'$ is updated and if its value 
    becomes lower than another point stored in memory, the anti-optimization process stops and the optimization
    continues to find a new optimum. A competition between optimization and anti-optimization is thus used to 
    enhance the efficiency of the robust algorithm by decreasing the number of points evaluated and the computational cost.

    To increase the reliability of the tolerance evaluation, the errors made simultaneously on several
    structure parameters have been taken into account by comparing the new computed design to the optimum
    known by the swarm. Since the particles move towards their local and global optimum, if a new minimum
    is computed around the optimum within the tolerance area $x^l_p \pm \Delta X$ or $x^g_p \pm \Delta X$ during
    the optimization process, its figure of merit $MF$ will be used to adjust the figure of merit $MF'$ associated
    to $x^l_p$ or $x^g_p$. The PSO exploration of the search space is thus used advantageously to increase the robustness
    by taking into account errors made simultaneously on several design parameters.

  \subsection{Tolerance requirements and first result}
    The execution of the robust optimization algorithm imposes to define tolerance requirements for the mirror parameters.
    One of the most critical parameter to control is the oxide layer thickness $T_A$ since a 8--13~\% decrease of the layer
    arises from the oxidation process of AlAs\cite{almuneau_jopt_2011}. The tolerance of the AlO$_x$ layer thickness has been
    set to be of $\Delta T_A = \pm$~50~nm during the optimization. Nevertheless, simulations have already shown that the low
    index sublayer thickness $T_A$ is not a critical parameter\cite{almuneau_jopt_2011} of the HCG mirror and a 50~nm tolerance
    should not be a severe design constraint\cite{chevallier_apa_2010,chevallier_oqe_2012}. The grating parameters $FF$ and $T_g$
    linked to the etching process are more critical and require large tolerance values of $\Delta T_g = \pm$~20~nm and $\Delta FF= \pm$0.02.
    The other parameters are either defined by the epitaxial growth of the structure ($T_L$) or by e-beam lithography ($\Lambda$) and are better controlled.     
    Tolerance requirements for these parameters have been set to lower values with $\Delta T_L = \pm$ 1~nm and $\Delta \Lambda = \pm$3~nm.

    The optimization of the GaAs/AlO$_x$ mirror with the robust optimization algorithm performed 
    in the same conditions as described in Section~\ref{part_Optim} results in a structure which dimensions
    are given in Table~\ref{table_RO_PSO} and exhibits a 369~nm large bandwidth. A tolerance study of the mirror
    parameters shows that all dimensions meet the tolerance requirements and that the optimum is correctly centered
    within the variation ranges (Table~\ref{table_RO_PSO}). A comparison with the design optimized with the particle 
    swarm algorithm without anti-optimization (Table~\ref{table_APSO}) shows that the bandwidth is 124~nm smaller for 
    the mirror obtained by robust optimization. A decrease of tolerance values is also exhibited for $\Delta T_g$ and 
    $\Delta FF$ but are still larger than the requirements. However, when performing a statistical test by varying simultaneously
    and randomly the structure dimensions with a uniform law for the previously defined tolerance values, the failure rate ($R_{TM} < 99.5$~\% at 
    $\lambda_0$) decreases from 1.9~\% from the design without robust optimization to less than 0.01~\% for the robust 
    design. Despite a decrease of bandwidth and tolerances, the robustness of the grating is still enhanced by the use 
    of the robust optimization algorithm which improves tolerance with respect to errors made on several parameters and
    guarantees the resulting optimized design to meet user-defined tolerance requirements. 

    \begin{table}[!t]
    \renewcommand{\arraystretch}{1.2}
    \caption{Optimum and tolerance values obtained by the robust optimization algorithm.}
    \label{table_RO_PSO}
    \centering
    \begin{tabular}{llll}
      \hline
      & Optimum & \multicolumn{2}{l}{Tolerances for $R_{TM} > 99.5\%$ at $\lambda_0$ }  \\
      \hline 
      $T_g$ &
	  713~nm &
	  682~nm $< T_g <$ 773~nm &
	  $\Delta T_g = \pm$31~nm
	  \\
      $FF$ &
	  0.484 &
	  0.403 $< FF <$ 0.548 &
	  $\Delta FF =  \pm$0.064
	  \\
      $T_A$ &
	  355~nm &
	  255~nm $< T_A $ & 
	  $\Delta T_A > \pm$100~nm
	  \\
      $\Lambda$ &
	1145~nm &
	1038~nm $< \Lambda <$ 1206~nm &
	  $\Delta \Lambda = \pm$61~nm
	  \\
      $T_L$ &
	  249~nm &
	  215~nm $< T_L <$ 301~nm &
	   $\Delta T_L = \pm$34~nm
	  \\
      \hline
      $\lambda_0$ &
	  \multicolumn{3}{l}{2309~nm }
	   \\
      $\Delta\lambda$ &
	  \multicolumn{3}{l}{369~nm }
	  \\
      $\Delta\lambda/\lambda_0$&
	  \multicolumn{3}{l}{16.0 \% }
	  
    \end{tabular}
    \end{table}

  \subsection{Final robust design}

    \begin{figure*}[!t]
      \centering
      \includegraphics[width=4.5in]{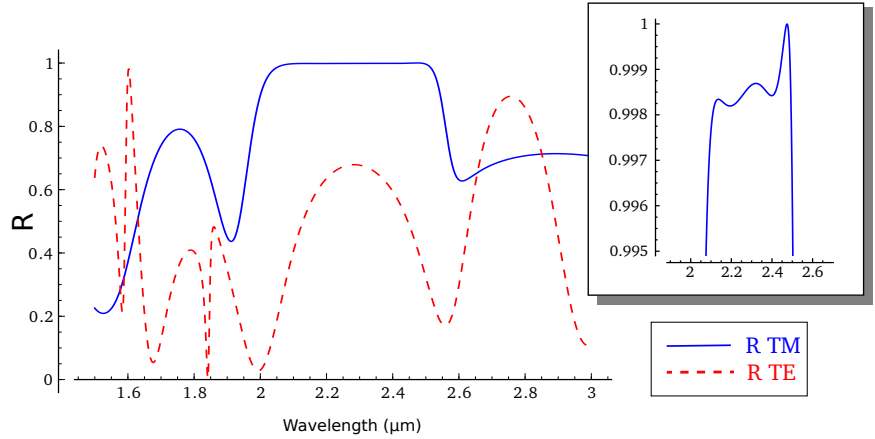}
      \caption{Reflection spectra of the robust HCG mirror designed with the parameters described in Table~\ref{table_RO_APSO_MT}.
      The inset exhibits a large 99.5\% high reflectivity bandwidth of 425~nm centered at 2290~nm for the TM coefficient (blue) with
      a good polarization selectivity by keeping $R_{TE} <$ 70~\% (dashed red).}
      \label{spectre_RO_APSO}
    \end{figure*}

    The use of a robust optimization algorithm allows the search of not only the most efficient but also robust
    solution. However, with a 25~\% decrease of the bandwidth and smaller tolerances for the critical parameters
    $T_g$ and $FF$, the optimum found by the robust optimization algorithm described in section~\ref{section_AO}
    seems to be a poor result compared to the non robust optimization. The combination of the anti-optimization 
    and the particle swarm algorithm has introduced a perturbation of the convergence of the optimization algorithm.
    When a new robust optimum is discovered, it could take several iterations of the algorithm to perform a good enough
    exploration of its neighborhood and decrease the robustness if a new minimum is found in the user defined tolerance ranges.
    A new optimum is then chosen by the swarm to be the best one, but its robustness has to be evaluated from scratch by the
    anti-optimization process and will not take into account the previously computed points. 

    To enhance the speed of the tolerance evaluation process of the anti-optimization, the robust optimization algorithm
    has been improved. Instead of using only the new points computed during the optimization to adjust the tolerance of 
    the best known design, the minimum used in the figure of merit $MF'$ is made on every points computed since the start
    of the optimization algorithm. This requires to store in memory all the tested designs and their associated figure of
    merit $MF$. The process of adding a memory in the swarm allows it to remember where the designs which are sensitive to
    error of fabrication are located and avoid them more easily.

    The execution of the modified robust optimization with the same technological and tolerance requirements as before
    returns a HCG mirror with a 425~nm large bandwidth which dimensions are given in Table~\ref{table_RO_APSO_MT}. 
    The tolerance evaluation of the parameters shows a good robustness with an optimum value well centered within 
    large variation range. The grating thickness $T_g$ and Fill Factor $FF$ tolerances are very large with $T_g = 668 \pm 70$~nm and $FF = 0.5351 \pm 0.0630$. 
    A statistic study with 30 000 tests of the tolerance of the optimum by varying simultaneously all the design parameters with 
    $\Delta T_g = \pm$~20~nm, $\Delta FF= \pm$0.02, $\Delta T_A = \pm$~50~nm, $\Delta \Lambda = \pm$~3~nm and  $\Delta T_L = \pm$~1~nm 
    have returned 0 mirror with less than 99.5~\% of TM reflectivity at $\lambda_0 = $~2300~nm.

    Finally, the tolerance of the optical index values of the materials have been explored. Indeed,
    the refractive index of the GaAs, and the AlOx especially, can vary with respect to the conditions of fabrication.
    Computations have shown that the high contrast grating presented in Table~\ref{table_RO_APSO_MT} exhibits a reflictivity
    of more than 99.5~\% at $\lambda_0$ for GaAs index values between 3.22 and 3.6 and AlOx index values between 1 and 1.78.
    Besides the variation of the refractive index, AlOx can present absorption in the mid infrared wavelength range~\cite{ravaro_apl_2008}.
    Even though the absorption is negligible below 2.5~\textmu m~\cite{ravaro_apl_2008}, the maximum absorption allowed to keep a 99.5~\% 
    reflectivity has been computed and is $\alpha =$~110~cm$^{-1}$ ($k = 0.00183$).

    \begin{table}[!t]
    \renewcommand{\arraystretch}{1.2}
    \caption{Optimum and tolerance values obtained by the enhanced robust optimization algorithm.}
    \label{table_RO_APSO_MT}
    \centering
    \begin{tabular}{lllll}
      \hline
      & Optimum & \multicolumn{2}{l}{Tolerances for $R_{TM} > 99.5\%$ at $\lambda_0$}  \\
      \hline 
      $T_g$ &
	  668~nm &
	  598~nm $< T_g <$ 738~nm &
	  $\Delta T_g = \pm$70~nm
	  \\
      $FF$ &
	  0.5351 &
	  0.4361 $< FF <$ 0.5981 &
	  $\Delta FF =  \pm$0.063
	  \\
      $T_A$ &
	  360~nm &
	  228~nm $< T_A $ & 
	  $\Delta T_A = \pm$132~nm
	  \\
      $\Lambda$ &
	1098~nm &
	993~nm $< \Lambda <$ 1162~nm &
	  $\Delta \Lambda = \pm$64~nm
	  \\
      $T_L$ &
	  282~nm &
	  240~nm $< T_L <$ 331~nm &
	   $\Delta T_L = \pm$42~nm
	  \\
      \hline
      $\lambda_0$ &
	  \multicolumn{3}{l}{2290~nm }
	   \\
      $\Delta\lambda$ &
	  \multicolumn{3}{l}{425~nm }
	  \\
      $\Delta\lambda/\lambda_0$&
	  \multicolumn{3}{l}{18.6 \% }
	  
    \end{tabular}
    \end{table}

\section{Conclusion}

A robust optimization algorithm has been developed to design high contrast grating mirrors 
for a VCSEL application at 2.3~\textmu m. An anti-optimization process based on particle swarm optimization
is used to adjust the geometrical parameters of the HCG structure by taking into account technological constraints.
The fabrication accuracy of our equipments is also taken into account within the optimization process by defining 
tolerance requirements that the HCG parameters have to meet. The execution of the robust optimization algorithm thus
results not only in an efficient mirror with a 99.5~\% high reflectivity for a 425~nm large bandwidth but also in a
robust design with more than $\pm$10~\% of tolerance on the grating thickness, which is one of the most critical 
parameter of the grating fabrication process. The mirror also exhibits a strong polarization selectivity by keeping
the reflection coefficient of the TE mode lower than 70~\%. This polarization selectivity combined with high mirror efficiency and 
large fabrication tolerance should make the GaAs/AlO$_x$ HCG design presented in this work a very good VCSEL mirror to allow emission above 2.3~\textmu m.

\section*{Acknowledgments}

The authors thank the French ANR for financial support in the framework of Marsupilami project (ANR-09-BLAN-0166-03) and 
IES and LAAS (France), partners of LMOPS/Sup\'elec in this project. This work was also  partly funded by the InterCell 
grant (http://intercell.metz.supelec.fr) by INRIA and R\'egion Lorraine (CPER2007).

\ifCLASSOPTIONcaptionsoff
  \newpage
\fi

\bibliographystyle{IEEEtran}

\end{document}